# Hierarchies of Inefficient Kernelizability

Danny Hermelin[1], Stefan Kratsch[2], Karolina Sołtys[3], Magnus Wahlström[1], and Xi Wu[4]

[1] Max Plank Institute for Informatics,
Stuhlsatzenhausweg 85, Saarbrücken 66123 - Germany
{hermelin,wahl}@mpi-inf.mpg.de
[2] Utrecht University,
Princetonplein 5, 3584 CC Utrecht - the Netherlands
s.kratsch@uu.nl
[3] Computer Science Division, University of California at Berkeley,
Soda Hall, Berkeley, CA 94720 - USA
ksoltys@cs.berkeley.edu
[4] Computer Science Department, University of Wisconsin Madison,
1210 W. Dayton St., Madison, WI 53706 - USA
xiwu@cs.wisc.edu

**Abstract.** The framework of Bodlaender et al. (ICALP 2008) and Fortnow and Santhanam (STOC 2008) allows us to exclude the existence of polynomial kernels for a range of problems under reasonable complexity-theoretical assumptions. However, there are also some issues that are not addressed by this framework, including the existence of Turing kernels such as the "kernelization" of LEAF OUT BRANCHING($k$) into a disjunction over $n$ instances of size poly($k$).

Observing that Turing kernels are preserved by polynomial parametric transformations, we define a *kernelization hardness* hierarchy, akin to the M- and W-hierarchy of ordinary parameterized complexity, by the PPT-closure of problems that seem likely to be fundamentally hard for efficient Turing kernelization. We find that several previously considered problems are complete for our fundamental hardness class, including MIN ONES $d$-SAT($k$), BINARY NDTM HALTING($k$), CONNECTED VERTEX COVER($k$), and CLIQUE($k \log n$), the clique problem parameterized by $k \log n$.

## 1 Introduction

Parameterized complexity theory [18, 21] is concerned with whether problems can be solved in $f(k) \cdot n^{O(1)}$ time, where $n$ is the total input size, $k$ a parameter typically independent of the input size, and $f()$ an arbitrary computable function. Such problems are called *fixed parameter tractable*, and FPT is the class of all fixed parameter tractable problems. In this sense, FPT extends the class P of tractable problems in classical complexity theory, allowing a refined analysis of hard computational problems. Complementing this notion of tractability is a set of classes of *fixed parameter intractability*, which allow classifying problems that are unlikely to be in FPT. These classes are organized into two main hierarchies, the W- and the M-hierarchy, which intertwine together to form an infinite hierarchy of intractability:

$$\text{FPT} \subseteq \text{M}[1] \subseteq \text{W}[1] \subseteq \text{M}[2] \subseteq \text{W}[2] \subseteq \cdots$$

Arguably the most useful technique in parameterized complexity is *kernelization*. A kernelization algorithm (or kernel) is a polynomial-time reduction from a problem to itself that compresses any problem instance to an equivalent instance of $f(k)$ size. Appropriately, the function $f()$ is referred to as the *size* of the kernel. Not only is kernelization one of the most successful techniques for showing that a problem is fixed-parameter tractable, it also provides an equivalent way of defining fixed-parameter tractability: A problem is solvable in $f(k) \cdot n^{O(1)}$ time iff it has a kernel [13]. Moreover, kernelization embodies within it the ubiquitous technique of preprocessing (data reduction), and gives the first natural and meaningful framework for analyzing this technique.

Since any fixed parameter tractable problem has a kernel, it is natural to ask which problems admit particularly efficient kernels, which have traditionally been defined as kernels with polynomial size bounds. Problems admitting polynomial kernels form a natural subclass of FPT, and in fact (in the case of problems in NP) also of EXPT, the class of parameterized problems solvable in $2^{k^{O(1)}} \cdot n^{O(1)}$ time. Examples of polynomial kernels are in abundance, including the linear kernels for VERTEX COVER [34] and PLANAR DOMINATING SET [2], the quadratic kernel for FEEDBACK VERTEX SET [35], and the meta-theorems for kernelization on bounded genus graphs [7] (see also the surveys in [5, 23]).

In recent years there has been increasing study in lower bounds for kernelization, and in particular in determining which problems are unlikely to admit polynomial kernels. This research effort started with the work of Bodlaender *et al.* [6], which developed a machinery for excluding polynomial size kernels under the assumption that the polynomial hierarchy (PH) does not collapse, relying on a key lemma of Fortnow and Santhanam [22]. This machinery was used to show that problems such as PATH($k$) and CLIQUE($w$) (the classical clique problem parameterized by the treewidth of the input graph) do not have polynomial-size kernels unless PH collapses [6]. Extensions of this framework were not late to appear [8, 10, 15–17], and were used to exclude polynomial kernels for numerous problems, including LEAF OUT BRANCHING($k$) [20], DISJOINT CYCLES($k$) [10], CONNECTED VERTEX COVER($k$) [17], and several CSP problems [29, 30], to name just a few.

The above mentioned lower bound mechanisms thus proved very useful in determining which problems allow polynomial kernelizations. However, there are relaxed, yet still interesting notions of efficient kernelization for which these frameworks do not apply. In particular, they do not allow the exclusion of polynomial Turing kernelizations, where the kernel reduction is of Turing-type rather than of Karp-type. This is not merely a theoretical notion. For example, the LEAF OUT BRANCHING($k$) problem, which as mentioned above has no polynomial kernel unless PH collapses, admits a kernel of size $O(k^3)$ as soon as the root of the out-branching has been selected [20]. As another example, one might consider the CLIQUE($\Delta$) problem, the classical clique problem parameterized by the maximum degree of the graph. Using *e.g.* [6], it is trivial to exclude the existence of a polynomial kernel for this problem, but simply making one initial selection of a vertex that is to be a member of the clique reduces the instance down to size $\Delta$. Thus, both these problems have Turing kernels where an instance is transformed into a disjunction over $n$ instances of small size. Clearly, in general one may also expect the existence of more involved Turing kernels for other problems.

To obtain lower bounds for Turing kernelizations, we adopt a different approach than previously considered. First of all, we observe that polynomial Turing kernels are preserved under so-called *polynomial parametric transformations* (PPTs), a type of reduction introduced by Bodlaender *et al.* [10] to exclude regular polynomial kernelizations. We then identify a hierarchy of problems in EXPT which we believe to be hard to kernelize, even under the relaxed notion of Turing reductions. These problems are reparameterizations of satisfiability problems used to define the W- and M-hierarchies. We then consider the hierarchies of classes defined by taking the PPT-closure of these problems, which leads to two hierarchies of classes: The WK-hierarchy and the MK-hierarchy. These hierarchies refine the class EXPT, intertwining together to form a tower of inclusions similar to the one formed by the W- and M-hierarchies:

$$\text{MK}[1] \subseteq \text{WK}[1] \subseteq \text{MK}[2] \subseteq \text{WK}[2] \subseteq \text{MK}[3] \subseteq \cdots \subseteq \text{EXPT} \subseteq \text{FPT}.$$

The class MK[1] corresponds to problems with polynomial kernels. Thus, the fundamental hardness class of our hierarchies is WK[1], which plays the same role for kernelizability as W[1] plays for FPT-time algorithms. Similarly to the expectation that FPT $\neq$ W[1], there are strong reasons to believe that WK[1]-hard problems do not admit any form of efficient kernelizations.

In particular, if any WK[1]-hard problem admits a polynomial Turing kernelization, then so do all problems in WK[1], including:

- The BINARY NDTM HALTING($k$) problem, which is the $k$-step halting problem for non-deterministic single-tape Turing machines with a binary tape alphabet, where $k$ is taken as the parameter. This problem is PPT-equivalent to NDTM HALTING($k \log n$), the $k$-step halting problem for general single-tape Turing machines under the parameter $k \log n$ (where $n$ is the total input size). A polynomial Turing kernel for any of these two problems implies polynomial Turing kernelizations for all problems in NP for which a witness can be verified (even non-deterministically) in time polynomial only in the witness size (as opposed to the total input size).
- MIN ONES $d$-SAT($k$), where the parameter is taken to be the Hamming weight $k$ of the solution, i.e. the number of variables set to true. Efficient kernelization for MIN ONES $d$-SAT($k$) implies efficient kernelization for all minimization problems for which consistency can be verified by local conditions, e.g. the $\mathcal{H}$-FREE EDGE MODIFICATION($k$) problem, where $\mathcal{H}$ is a finite set of forbidden induced subgraphs, and the goal is to remove or add $k$ edges in the input graph in order to obtain a graph with no induced subgraph in $\mathcal{H}$ [12]. Note that some of these problems are not immediately covered by the previous item.
- Most natural W[1]-complete problems reparameterized under the parameter $k \log n$. For example, the ubiquitous clique problem is WK[1]-complete problem under this parameterization. If one believes in the fundamental hardness of the clique problem, then it seems unlikely that a polynomial-time Turing reduction would reduce it to questions about graphs of size $(k \log n)^{O(1)}$.
- Several EXPT problems under the standard natural parameterization of solution size, which turn out to be complete for WK[1]. These include UNIQUE COVERAGE($k$), MULTICOLORED PATH($k$), and CONNECTED VERTEX COVER($k$).

The remainder of the paper is organized as follows. In Section 2 we give precise definitions for the central concepts used in this paper. Section 3 is then used to define our classes of inefficient kernelizability, namely the WK- and MK-hierarchies, and also to discuss basic properties of these hierarchies. The main technical body of our work is presented in Section 4, where we prove that several problems are complete for our fundamental hardness class WK[1], while in Section 5 we discuss problems that reside in higher levels of our hierarchies. We conclude the paper in Section 6 by posing some open problems.

## 2 Preliminaries

We begin our discussion by formally defining some of the main concepts used in this paper, and by introducing some terminology and notation that will be used throughout. We use $[n]$ to denote the set of integers $\{1, \ldots, n\}$ for any integer $n \geq 1$.

**Definition 1 (Kernelization).** *A* kernelization algorithm, *or, in short, a* kernel *for a parameterized problem $L \subseteq \Sigma^* \times \mathbb{N}$ is a polynomial-time algorithm that on a given input $(x, k) \in \Sigma^* \times \mathbb{N}$ outputs a pair $(x', k') \in \Sigma^* \times \mathbb{N}$ such that*

- *$(x, k) \in L \Leftrightarrow (x', k') \in L$, and*
- *$|x'| + k' \leq f(k)$ for some function $f()$.*

*The function $f()$ above is referred to as the* size *of the kernel.*

In other words, a kernel is a polynomial-time reduction from a problem to itself that compresses the problem instance to a size depending only on the parameter. If the size of a kernel for $L$ is polynomial, we say that $L$ has a *polynomial kernel*. In the interest of robustness and ease of presentation, we relax the notion of kernelization to also allow the output to be an instance of a different problem. This has been referred to as a generalized kernelization [6] or bikernelization [3]. The class of all parameterized problems with polynomial kernels in this relaxed sense is denoted by PK.

**Definition 2 (Turing Kernelization).** *A* Turing kernelization *for a parameterized problem $L \subseteq \Sigma^* \times \mathbb{N}$ is a polynomial-time algorithm with oracle access to a parameterized problem $L'$ that can decide whether an input $(x, k)$ is in $L$ using queries of size bounded by $f(k)$, for some computable function $f()$. Again, the function $f()$ is referred to as the* size *of the kernel.*

If the size is polynomial, we say that $L$ has a *polynomial Turing kernel*. The class of all parameterized problems with polynomial Turing kernels is denoted by Turing-PK.

**Definition 3 (Polynomial Parametric Transformations [10]).** *Let $L_1$ and $L_2$ be two parameterized problems. We write $L_1 \leq_{ppt} L_2$ if there exists a polynomial time computable function $\Psi : \{0,1\}^* \times \mathbb{N} \to \{0,1\}^* \times \mathbb{N}$ and a constant $c \in \mathbb{N}$, such that for all $(x, k) \in \Sigma^* \times \mathbb{N}$, if $(x', k') = \Psi(x, k)$ then:*

- *$(x, k) \in L_1 \iff (x', k') \in L_2$, and*
- *$k' \leq ck^c$.*

*The function $\Psi$ is called a* polynomial parameter transformation *(PPT for short). If $L_1 \leq_{ppt} L_2$ and $L_2 \leq_{ppt} L_1$ we write $L_1 \equiv_{ppt} L_2$.*

**Lemma 1.** *Let $L_1$, $L_2$, and $L_3$ be three parameterized problems.*

- *If $L_1 \leq_{ppt} L_2$ and $L_2 \leq_{ppt} L_3$ then $L_1 \leq_{ppt} L_3$.*
- *If $L_1 \leq_{ppt} L_2$ and $L_2 \in$ PK (resp. Turing-PK) then $L_1 \in$ PK (resp. Turing-PK).*

For $t \geq 0$ and $d \geq 1$, we inductively define the following classes $\Gamma_{t,d}$ and $\Delta_{t,d}$ of formulas following [21]:

$$\begin{aligned}
\Gamma_{0,d} &:= \{\lambda_1 \wedge \cdots \wedge \lambda_c : c \in [d] \text{ and } \lambda_1, \ldots, \lambda_c \text{ are literals }\}, \\
\Delta_{0,d} &:= \{\lambda_1 \vee \cdots \vee \lambda_c : c \in [d] \text{ and } \lambda_1, \ldots, \lambda_c \text{ are literals }\}, \\
\Gamma_{t+1,d} &:= \{\bigwedge_{i \in I} \delta_i : I \text{ is a finite non-empty index set and } \delta_i \in \Delta_{t,d} \text{ for all } i \in I\}, \\
\Delta_{t+1,d} &:= \{\bigvee_{i \in I} \gamma_i : I \text{ is a finite non-empty index set and } \gamma_i \in \Gamma_{t,d} \text{ for all } i \in I\}.
\end{aligned}$$

Thus, $\Gamma_{1,3}$ is the set of all 3-CNF formulas, and $\Gamma_{2,1}$ is the set of all CNF formulas. Given a class $\Phi$ of propositional formulas, we let $\Phi^+, \Phi^- \subseteq \Phi$ denote the restrictions of $\Phi$ to formulas containing only positive and negative literals, respectively. For any given $\Phi$, we define two parameterized problems:

- $\Phi$-WSAT$(k \log n)$ is the problem of determining whether a formula $\phi \in \Phi$ with $n$ variables has a satisfying assignment of Hamming weight $k$, parameterized by $k \log n$.
- $\Phi$-SAT$(n)$ is the problem of determining whether a formula $\phi \in \Phi$ with $n$ variables is satisfiable, parameterized by $n$.

In particular, we will be interested in $\Gamma_{t,d}$-WSAT$(k \log n)$ and $\Gamma_{t,d}$-SAT$(n)$.

## 3 The WK- and MK-Hierarchies

In the following section we introduce our hierarchies of inefficient kernelizability, the MK- and WK-hierarchies. For a parameterized problem $L \subseteq \Sigma^* \times \mathbb{N}$, we let $[L]_{\leq_{ppt}}$ denote the closure of $L$ under polynomial parametric transformations. That is, $[L]_{\leq_{ppt}} := \{L' \subseteq \Sigma^* \times \mathbb{N} : L' \leq_{ppt} L\}$.

**Definition 4.** *Let $t \geq 1$ be an integer. The classes* WK[$t$] *and* MK[$t$] *are defined by*

- WK[$t$] := $\bigcup_{d \in \mathbb{N}} [\Gamma_{t,d}\text{-WSAT}(k \log n)]_{\leq_{ppt}}$.
- MK[$t$] := $\bigcup_{d \in \mathbb{N}} [\Gamma_{t,d}\text{-SAT}(n)]_{\leq_{ppt}}$.

The naming of the classes in our hierarchies comes from the close relationship of the MK- and WK-hierarchies to the M- and W-hierarchies of traditional parameterized complexity [21]. Roughly speaking, WK[$t$] and MK[$t$] are reparameterizations by a factor of $\log n$ of the traditional parameterized complexity classes W[$t$] and M[$t$] (although W[$t$] and M[$t$] are closed under FPT reductions, which may use superpolynomial time in $k$). There are also close connections to the so-called *subexponential time* S-hierarchy (see [21, Chapter 16]); specifically, S[$t$] and MK[$t$] are defined from the same starting problems, using closures under different types of reduction.

Yet another related hierarchy is the *VC-hierarchy* defined by Harnik and Naor [24]. The VC-hierarchy concerns a notion of *instance compression* which Harnik and Naor argue to be similar to but distinct from polynomial kernelization. Without going into technical details, we note that the problem LOCAL CIRCUIT SAT, which defines their class VC$_1$, is WK[1]-complete under a parameter of $k \log n$ (see next section), while the defining problem for VC$_t$ for $t > 1$ is trivially MK[$t$]-complete under a parameter of $n$. See [24] for all definitions.

**Theorem 1.** *Let $t \geq 1$. The following hold.*

- $\Gamma_{1,2}^-$*-WSAT$(k \log n)$ is* WK[1]*-complete.*
- $\Gamma_{t,1}^-$*-WSAT$(k \log n)$ is* WK[$t$]*-complete for odd $t > 1$.*
- $\Gamma_{t,1}^+$*-WSAT$(k \log n)$ is* WK[$t$]*-complete for even $t > 1$.*
- $\Gamma_{1,d}$*-SAT$(n)$ is* MK[1]*-complete for every $d \geq 3$.*
- $\Gamma_{t,1}$*-SAT$(n)$ is* MK[$t$]*-complete for $t \geq 2$.*

Theorem 1 above shows that the traditional problems used for showing completeness in the W- and M-hierarchies have reparameterized counterparts which are complete for our hierarchy. The theorem is proven using a set of PPTs from the specific class-defining problems to the corresponding target problem in the theorem. For the first item in the theorem, previous proofs have used FPT-time reductions; we provide a PPT. The remaining items are either easy or well-known.

**Lemma 2.** *Let $d \geq 1$. Then $\Gamma_{1,d}$-WSAT$(k \log n) \leq_{ppt} \Gamma_{1,2}^-$-WSAT$(k \log n)$.*

*Proof.* The lemma is trivial for $d = 1$, so assume $d \geq 2$. We show the proof in four steps. First we transform our input formula into an anti-monotone $\Gamma_{1,d}^-$ formula. The anti-monotone formula is then transformed into a multicolored $\Gamma_{1,d}^-$ formula, then to a multicolored $\Gamma_{1,2}^-$ formula, and finally to an uncolored $\Gamma_{1,2}^-$ formula. Since the transformation at each step will be a PPT, this will prove the lemma.

$\Gamma_{1,d}$-WSAT$(k \log n) \leq_{ppt} \Gamma_{1,d}^-$-WSAT$(k \log n)$: To transform to an anti-monotone formula, we adapt a trick used in [21]. Let $\phi$ be a $\Gamma_{1,d}$-formula on variables $X = \{x_1, \ldots, x_n\}$. We introduce a new set of variables $y_{i,j,j'}$, with the interpretation "the $i$'th true variable is $x_j$ and the $(i+1)$'th true variable is $x_{j'}$", taken over the ordering of $X$ assumed above. We will convert the formula $\phi$ into a formula $\phi'$ using only the variables $Y$. By a combination of the Hamming weight condition and negative 2-clauses, we can enforce consistency among the $y$-variables. Specifically, it is easy to enforce the following structure on the solutions of the formula:

- At most one $y$-variable is true for every $1 \leq i < k$.
- If $y_{i,j_1,j_2}$ and $y_{i+1,j_3,j_4}$ are both true, then $j_1 < j_2 = j_3 < j_4$.
- If $y_{i,j_1,j_2}$ and $y_{i',j_3,j_4}$ are both true, and $i' > i+1$, then $j_1 < j_2 < j_3 < j_4$.

Thus, if there are $k-1$ true $y$-variables, then these correspond to an ordered sequence of $k$ variables $x_j$. Let $\phi''$ be the formula containing these clauses. We can now replace every clause of $\phi$ by a conjunction of negative clauses, as follows. First, we may replace every literal $x_j$ or $\neg x_j$ by a conjunction of negative literals of $y$-variables:

$$x_j = \bigwedge_{j_1 > j, j_2 > j} \neg y_{1,j_1,j_2} \wedge \bigwedge_{i, j_1 < j, j_2 > j} \neg y_{i,j_1,j_2} \wedge \bigwedge_{j_1 < j, j_2 < j} \neg y_{k-1,j_1,j_2}$$

$$\neg x_j = \bigwedge_{i,j'} \neg y_{i,j,j'} \wedge \bigwedge_{j'} \neg y_{k-1,j',j}$$

To see why these equalities hold, negate both sides of the equations; the result is an obvious description of $x_j$ and $\neg x_j$ as disjunctions over positive $y$-variables. Second, since every clause in $\phi$ has bounded size, we may multiply out these conjunctions (using the distributive law of conjunctions) into individual $d$-clauses over literals $\neg y_{i,j,j'}$. Let $\phi'$ be the resulting formula. Then $\phi' \wedge \phi''$ is a $\Gamma^-_{1,d}$-formula and has a satisfying assignment of weight $k' = k-1$ if and only if $\phi$ has a satisfying assignment of weight $k$.

$\Gamma^-_{1,d}$-WSAT($k \log n$) $\leq_{ppt}$ Multicolored $\Gamma^-_{1,d}$-WSAT($k \log n$): Let $(\phi, k)$ be an instance of $\Gamma^-_{1,d}$-WSAT($k \log n$). We will produce an equivalent multicolored instance, where variables come in one of $k$ colors, and the solution is required to contain exactly one variable of each color. This is easy since $\phi$ is anti-monotone: Create $k$ copies of the entire variable set, each colored in a different color. For each variable $x_i$ in $\phi$, let $x_{i,c}$ be the copy of $x_i$ of color $c$. For every $x_i$ and every pair of colors $c \neq c'$, add a clause $(\neg x_{i,c} \vee \neg x_{i,c'})$, ensuring that $k$ distinct variables are chosen. Then, for every clause $(\neg x_1 \vee \ldots \vee \neg x_d)$ in $\phi$, replace it by one clause $(\neg x_{1,c_1} \vee \ldots \vee \neg x_{d,c_d})$ for every set of distinct colors $c_1, \ldots, c_d \in [k]$. Let $\phi'$ be the resulting formula. Note that each clause in $\phi$ excludes only a specific combination $x_1 \wedge \ldots \wedge x_d$, while the set of replacing clauses in $\phi'$ collectively excludes every possible selection of $x_1, \ldots, x_d$ from different colors; thus $\phi'$ has a multicolored satisfying assignment of weight $k$ iff $\phi$ has a satisfying assignment of weight $k$.

Multicolored $\Gamma^-_{1,d}$-WSAT($k \log n$) $\leq_{ppt}$ Multicolored $\Gamma^-_{1,2}$-WSAT($k \log n$): Let $\phi$ be a multicolored $\Gamma^-_{1,d}$-formula on variables $X$, $|X| = n$, and $k$ colors. We create a multicolored $\Gamma^-_{1,2}$-formula $\phi'$ as follows: Let the colors of $\phi'$ correspond to $d$-tuples of colors from $\phi$; thus $\phi'$ has $k' = O(k^d)$ colors. For every color $C$ of $\phi'$, corresponding to colors $(c_1, \ldots, c_d)$ of $\phi$, create a set of variables as follows. Initialize with the set $X_C = \{(x_1, \ldots, x_d) : x_i \in X, x_i \text{ has color } c_i, 1 \leq i \leq d\}$. Then, remove from $X_C$ every tuple $(x_1, \ldots, x_d)$ which explicitly falsifies a clause of $\phi$. Since $\phi$ is anti-monotone, this is simple, e.g., if all clauses of $\phi$ are $d$-ary, then $(x_1, \ldots, x_d)$ is removed from $X_C$ if and only if $\phi$ contains a clause $(\neg x_1 \vee \ldots \vee \neg x_d)$. Let the remaining set of tuples be $X_C^*$. Now, for every pair of color-tuples $C$ and $C'$ which have at least one color in common, enforce consistency by excluding pairs of variable-tuples which disagree on some common color, using a conjunction of clauses with two negative literals each. Let the resulting instance be $\phi'$. It is clear that a multicolored satisfying assignment to $\phi'$ corresponds to a multicolored satisfying assignment for $\phi$. Furthermore, since $\phi$ is anti-monotone, every clause has been "verified" in the construction of some set $X_C^*$. Thus $\phi'$ has a multicolored satisfying assignment of weight $k'$ iff $\phi$ has a multicolored satisfying assignment of weight $k$.

Multicolored $\Gamma^-_{1,2}$-WSAT($k \log n$) $\leq_{ppt}$ $\Gamma^-_{1,2}$-WSAT($k \log n$): Given a multicolored $\Gamma^-_{1,2}$ formula $\phi$, we transform $\phi$ into an "uncolored" formula $\phi'$ by adding a conjunction of negative 2-clauses over every color class of $\phi'$. These ensure that exactly one variable of each class is set to true in any satisfying assignment of $\phi'$. Thus, $\phi'$ has a weight $k$ satisfying assignment iff $\phi'$ has a weight $k$ satisfying assignment. □

The following follows from Flum and Grohe [21]. It can be readily verified that every reduction used to prove [21, Lemma 7.5] is a PPT.

**Lemma 3 ([21, Lemma 7.5]).** *Let $d \geq 1$. Then:*

- $\Gamma_{t,d}\text{-WSAT}(k \log n) \equiv_{ppt} \Gamma_{t,1}^{-}\text{-WSAT}(k \log n)$ *for all odd $t > 1$.*
- $\Gamma_{t,d}\text{-WSAT}(k \log n) \equiv_{ppt} \Gamma_{t,1}^{+}\text{-WSAT}(k \log n)$ *for all even $t > 1$.*

Next, we give the reductions for the MK-hierarchy.

**Lemma 4.** *Let $d \geq 3$. Then $\Gamma_{1,d}\text{-SAT}(n) \leq_{ppt} \Gamma_{1,3}\text{-SAT}(n)$.*

*Proof.* Let $\phi$ be a $\Gamma_{1,d}$ formula, *i.e.* a $d$-CNF formula. By removing duplicate clauses, the total length of $\phi$ is at most $O(n^d)$. Since $d$ is a constant, the classical reduction to 3-CNF of Karp [28] (via clause splitting) is a PPT. □

**Lemma 5.** *Let $t \geq 2$ and $d \geq 1$. Then $\Gamma_{t,d}\text{-SAT}(n) \leq_{ppt} \Gamma_{t,1}\text{-SAT}(n)$.*

*Proof.* Let $\phi$ be a $\Gamma_{t,d}$-formula. Introduce a new variable for every bottom-level disjunction or conjunction in $\phi$ of arity at most $d$. It is simple to enforce the values of the newly introduced variables using a $\Gamma_{2,1}$-formula. Let $\phi'$ be such a formula, and let $\phi''$ be $\phi$ with every bounded-arity bottom-level conjunction or disjunction replaced by the corresponding new variable. Then $\phi \equiv \phi' \wedge \phi''$ and the reduction creates at most $O(n^d)$ variables. □

This finishes the set of reductions needed to prove Theorem 1. We now proceed to show the class containments in our hierarchy. Let us denote by $\widetilde{\text{NP}}$ the class of such parameterized problems whose unparameterized variants are in NP. That is, for a parameterized problem $L \subseteq \Sigma^* \times \mathbb{N}$, we let $\widetilde{L} \subseteq (\Sigma \cup \{\#\})^*$ denote the *unparameterized variant* of $L$ defined by $\widetilde{L} := \{x\#1^k : (x,k) \in L\}$, where $\#$ is some symbol not in $\Sigma$. Then $\widetilde{\text{NP}} = \{L \subseteq \Sigma^* \times \mathbb{N} : \widetilde{L} \in \text{NP}\}$. We have the following relationship between PK and MK[1]:

**Lemma 6.** $\text{PK} \cap \widetilde{\text{NP}} = \text{MK}[1]$.

*Proof.* Let $L$ be a parameterized problem in $\text{PK} \cap \widetilde{\text{NP}}$. Then $L$ admits a polynomial kernel $\mathcal{K}$, and since $\widetilde{L} \in \text{NP}$, there is a reduction $\mathcal{R}$ from $\widetilde{L}$ to the unparameterized version of $\Gamma_{1,3}\text{-SAT}(n)$. Composing $\mathcal{K}$ and $\mathcal{R}$ (by appropriately switching back and forth from instances of $L$ to instances of $\widetilde{L}$) gives a polynomial parameteric transformation from $L$ to $\Gamma_{1,3}\text{-SAT}(n)$. Thus, $L \in \text{MK}[1]$, and so $\text{PK} \cap \widetilde{\text{NP}} \subseteq \text{MK}[1]$. Conversely, an instance of $\Gamma_{1,d}\text{-SAT}(n)$ has $O(n^d)$ distinct clauses, and so by removing duplicate clauses we get a polynomial kernel for $\Gamma_{1,d}\text{-SAT}(n)$. Thus, $\Gamma_{1,d}(n) \in \text{PK}$ for every $d \geq 1$, and by Lemma 1 we have $\text{MK}[1] \subseteq \text{PK}$. Furthermore, as MK[1] is defined as the PPT-closure of problems whose unparameterized problems are in NP, and since PPTs are a restricted form of classical polynomial-time reductions, we have $\text{MK}[1] \subseteq \widetilde{\text{NP}}$. □

**Lemma 7.** $\text{MK}[t] \subseteq \text{WK}[t]$ *for any integer $t \geq 1$.*

*Proof.* Let $t \geq 1$. By Lemma 1 and Definition 4, to prove the lemma it suffices to show that $\Gamma_{t,d}\text{-SAT}(n) \leq_{ppt} \Gamma_{t,d+1}\text{-WSAT}(k \log n)$ for every $d \in \mathbb{N}$. So fix $d$, and let $\alpha$ be an $\Gamma_{t,d}$ formula on $n$ variables given as an input to $\Gamma_{t,d}\text{-SAT}(n)$. Let $x_1, \ldots, x_n$ denote the variables of $\alpha$. We construct a formula over the variables $x_1^0, x_1^1, \ldots, x_n^0, x_n^1$. Let $\alpha'$ denote the formula obtained after replacing in $\alpha$ each positive occurrence of $x_i$ with $x_i^1$, and each negative occurrence of $x_i$ with $x_i^0$. Then $\alpha' \in \Gamma_{t,d}$. For each $i \in [n]$, define $\beta_i$ to be the formula $\beta_i := (x_i^0 \vee x_i^1) \wedge (\neg x_i^0 \vee \neg x_i^1)$, i.e., $\beta_i \equiv x_i^0 \neq x_i^1$, and consider the formula $\alpha' \wedge \left(\bigwedge_i \beta_i\right)$. Then, as $\alpha' \in \Gamma_{t,d}$ and $\beta_i \in \Gamma_{1,2}$ for all $i \in [n]$, this formula can be written as a $\Gamma_{t,d'}$ formula $\beta$, where $d' = \max\{d, 2\} \leq d + 1$. Moreover, $\beta$ has a satisfying assignment of weight weight $k := n$ iff $\alpha$ is satisfiable. The lemma follows. □

**Lemma 8.** WK[$t$] ⊆ MK[$t + 1$] *for any integer $t \geq 1$.*

*Proof.* The lemma can be shown using the "$k \log n$ trick" introduced by Abrahamson *et al.* [1]. Introduce $k$ groups of $\log n$ variables, each group determining via binary expansion the identity of one selected variable. Since MK[$t+1$] may use formulas of depth one larger than WK[$t$], it is possible to essentially replace the literals in an input formula by expressions over the $k \log n$ new variables. See [21, Theorem 16.42] for a detailed construction. □

**Lemma 9.** MK[$t$], WK[$t$] ⊆ EXPT *for all $t \geq 1$.*

*Proof.* Note that membership in EXPT is preserved by a PPT. Thus by Lemma 8 it suffices that $\Gamma_{t,d}$-SAT($n$) ∈ EXPT for every $t$ and $d$, which is trivial by brute force. □

Combining Lemma 7, Lemma 8, and Lemma 9 we get the following tower of inclusions of hierarchies, refining the class EXPT.

**Theorem 2.** (PK ∩ $\widetilde{\text{NP}}$) = MK[1] ⊆ WK[1] ⊆ MK[2] ⊆ WK[2] ⊆ MK[3] ⊆ ⋯ ⊆ EXPT.

As we will see in the next section, we know that WK[1] ⊈ PK unless the polynomial hierarchy collapses, since several complete problems for WK[1] were previously shown not to have polynomial kernels under this assumption. We conjecture that none of these have polynomial Turing kernels as well. Thus, the main working hypothesis which this paper evolves around is the following:

**Conjecture:** WK[1] ⊈ Turing-PK.

## 4 Complete Problems for WK[1]

In the following section we exemplify the robustness of WK[1], the fundamental hardness class of our hierarchy. We do this by showing that several natural problems are complete for this class. In doing so, we also establish that these problems are unlikely to admit polynomial Turing kernels. We begin in Section 4.1, by showing WK[1]-completeness for a handful of fundamental problems which will be used later in the reductions for other problems. Following this, in Section 4.2 we study two important satisfiability problems. Then, in Section 4.3, we review some classical path and cycle problems which were shown in [10] not to admit polynomial kernels unless PH collapses. Following this, in Section 4.4, we examine problems whose kernelizabilty status has been dealt with in [17]. For the sake of brevity, we defer all standard problem definitions to Appendix A.

### 4.1 Basic Problems

In this section we prove the following theorem, which covers some basic problems which will be convenient for showing WK[1]-hardness and completeness for other problems.

**Theorem 3.** *The following problems are all complete for* WK[1]*:*

- CLIQUE($k \log n$) and INDEPENDENT SET($k \log n$).
- BINARY NDTM HALTING($k$) and NDTM HALTING($k \log n$).
- HITTING SET($m$) and EXACT HITTING SET($m$).
- SET COVER($n$) and EXACT SET COVER($n$).

Note that the result for CLIQUE($k \log n$) and INDEPENDENT SET($k \log n$) follows directly from the fact that both problems are known to PPT-equivalent to $\Gamma^-_{1,2}$-WSAT($k \log n$) (see e.g. [21]), and from the fact that $\Gamma^-_{1,2}$-WSAT($k \log n$) is WK[1]-complete (Lemma 3). We will also need a few annotated problems to be used in reductions. The following lemma may be considered folklore, although an explicit proof of its first part can be found in [19], and of the second part in [17][5].

**Lemma 10.** *The following equivalences hold.*

1. MULTICOLORED CLIQUE($k \log n$) $\equiv_{ppt}$ CLIQUE($k \log n$).
2. MULTICOLORED HITTING SET($m$) $\equiv_{ppt}$ HITTING SET($m$).

We now proceed with the reductions. For many problems, we will find it most convenient to show hardness by reduction from EXACT HITTING SET($m$) or HITTING SET($m$), and membership by reduction to a Turing machine problem; hence we begin by showing the completeness of these problems. We give a chain of reductions from MULTICOLORED CLIQUE($k \log n$) to EXACT HITTING SET($m$) to NDTM HALTING($k \log n$) to BINARY NDTM HALTING($k$), for which we finally show WK[1]-membership directly, closing the cycle; after this we will treat the HITTING SET($m$) problem.

**Lemma 11.** MULTICOLORED CLIQUE($k \log n$) $\leq_{ppt}$ EXACT HITTING SET($m$).

*Proof.* Let $G$ be a graph on $n$ vertices in an instance of MULTICOLORED CLIQUE($k \log n$), and let $c : V(G) \to [k]$ the coloring function of $G$. We assume that $V(G) = [n]$, and we let $b_\ell(v)$ denote the $\ell$'th bit in the binary expansion of $v \in V(G)$. The instance $(U, \mathcal{F})$ of EXACT HITTING SET($m$) is constructed by taking $U = E(G)$ and defining $\mathcal{F}$ to be the following subsets of $U$:

- $F_{i,j} := \{uv \in E(G) : c(u) = i, c(v) = j\}$ for all $1 \leq i < j \leq k$.
- $F_{i,j,j',\ell} := \{uv \in E(G) : c(u) = i, c(v) = j, b_\ell(u) = 1\} \cup \{uv \in E(G) : c(u) = i, c(v) = j', b_\ell(u) = 0\}$ for every pair of color pairs $(i, j)$ and $(i, j')$ with $j \neq j'$, and all $1 \leq \ell \leq \log n$.

Observe that any exact hitting set $U^* \subseteq U$ for $(U, \mathcal{F})$ consists of $\binom{k}{2}$ edges, one for each pair of distinct colors $i, j \in [k]$. The sets $F_{i,j,j',\ell}$, $1 \leq \ell \leq \log n$, ensure that any pair of edges $e \neq e' \in U^*$ with one endpoint colored $i$ are incident with the same vertex $v \in V(G)$ with $c(v) = i$. Otherwise, if $e$ is incident with an $i$-colored vertex $v$, and $e'$ is incident with an $i$-colored vertex $v' \neq v$, then $b_\ell(v) \neq b_\ell(v')$ for some $\ell \in \{1, \ldots, \log n\}$, and $e$ and $e'$ are both in some $F_{i,j,j',\ell}$ for this specific $\ell$. Thus, any exact hitting set of $(U, \mathcal{F})$ corresponds to a multicolored clique in $G$. Conversely, the edge-set of any multicolored clique in $G$ is an exact hitting set of $(U, \mathcal{F})$. As $m := |\mathcal{S}| = O(k^2 + k \log n)$, our construction is a polynomial parameteric transformation, and so MULTICOLORED CLIQUE($k \log n$) $\leq_{ppt}$ EXACT HITTING SET($m$). □

**Lemma 12.** EXACT HITTING SET($m$) $\leq_{ppt}$ NDTM HALTING($k \log n$).

---
[5] Dom *et al.* [17] actually use the name RED BLUE DOMINATING SET instead of HITTING SET, but the two problems are essentially the same.

*Proof.* Let $(U, \mathcal{F})$ be an instance of EXACT HITTING SET($m$), with $U = [n]$ and $\mathcal{F} = \{F_1, \ldots, F_m\}$. By identifying vertices which are included in the same set of edges, we may assume that $n \leq 2^m$. We create a Turing machine $M$ with alphabet $[n]$, which writes down $m$ values specifying the selected member of each set, followed by simple poly($m$)-time verification. Specifically, let the $m$ selections be $u_1, \ldots, u_m$. Then the machine has to verify that $u_i \in F_i$ for each $i \in [m]$, and that for every pair $i, j \in [m]$, either $u_i = u_j$ or $u_i \in (F_i \setminus F_j)$ and $u_j \in (F_j \setminus F_i)$. By encoding this information in the state space of $M$, we get a machine of size $n'$ which is polynomial in $n + m$, that can nondeterministically verify in $k = O(m^2)$ steps whether $(U, \mathcal{F})$ has an exact hitting set. Since $\log n' = O(m)$, this is indeed a PPT. □

**Lemma 13.** NDTM HALTING($k \log n$) $\leq_{ppt}$ BINARY NDTM HALTING($k$), *and both are members of* WK[1].

*Proof.* The reduction from NDTM HALTING($k \log n$) to BINARY NDTM HALTING($k$) is well-known: The cells of the tape of the former machine $M$, with entries from $\Sigma$, can be subdivided into $\log |\Sigma| \leq \log |M|$ binary cells, and every transition of the machine subdivided into $\log |\Sigma|$ steps corresponding to reading or writing the value of the original cell. Thus, it remains to show membership for BINARY NDTM HALTING($k$).

This we do by a reduction to $\Gamma_{1,3}$-WSAT($k \log n$). Let $(M, k)$ be the input to BINARY NDTM HALTING($k$), where $M$ is a Turing machine of size $n$ with $s$ states and $\ell$ edges in its transition diagram. We construct a $\Gamma_{1,3}$ (3-CNF) formula $\phi$, such that $\phi$ has a satisfying assignment of Hamming weight $k'$, value to be chosen later, iff $M$ accepts the empty string in $k$ steps. For this we create variable groups as follows:

- Variables $S_{i,t}$, $i \in [s]$ and $t \in \{0\} \cup [k]$, signifying that the machine is in state number $i$ after $t$ time steps.
- Variables $M_{e,t}$, $e \in [\ell]$ and $t \in [k]$, signifying that edge $e$ of the machine's state diagram is followed as step $t$.
- Variables $H_{p,t}$, $0 \leq p, t \leq k$, signifying that the machine's tape head is in position $p$ after $t$ time steps.
- Variables $T_{p,t}$, $0 \leq p, t \leq k$, signifying that position $p$ of the tape contains value 1 after $t$ time steps.
- Variables $\overline{T}_{p,t}$, $0 \leq p, t \leq k$. These are constrained so that $T_{p,t} \neq \overline{T}_{p,t}$ for all values of $p$ and $t$, allowing us to control the weight of any satisfying assignment for $\phi$.

We add the following clauses to $\phi$. We first add clauses of size 2 to ensure that $T_{p,t} \neq \overline{T}_{p,t}$ for every value of $p$ and $t$. Then we enforce that at most one variable among $S_{\cdot,t}$, $H_{\cdot,t}$, and $M_{\cdot,t}$ is true for each $t$, by creating a conjunction of clauses of size 2 over the corresponding variable sets. We additionally enforce, by negative clauses of size 1, that the final state of the machine is an accepting state. We then add clauses enforcing consistency of states, head, tape, and transition variables. For example, if edge $e$ moves from state $i$ to state $j$, reading a 0, writing a 1, and moving right, then we have constraints: $(M_{e,t} \wedge H_{p,t-1} \to \neg T_{p,t-1})$, $(M_{e,t} \wedge H_{p,t-1} \to T_{p,t})$, $(M_{e,t} \wedge H_{p,t-1} \to H_{p+1,t})$, $(M_{e,t} \to S_{i,t-1})$, and $(M_{e,t} \to S_{j,t})$, for all possible values of $p$ and $t$. We complete the construction of $\phi$ by adding clauses encoding $(\neg H_{p,t} \to (T_{p,t} = T_{p,t+1}))$ for all $p$ and $t$, and clauses controlling the initial setup of the machine (*e.g.* head position 0, state 1, all tape entries 0).

As a final step of our reduction we set $k' = 2(k+1) + (k+1)^2 + k$. This accounts for the fact that exactly one variable among the $S$-, $H$-, and $M$-variables is set to true, and that the total Hamming weight of the $T$- and $\overline{T}$-variables is $(k+1)^2$, which is the total number of bits needed to encode the information on the tape of $M$ throughout $k$ computation steps. The formula $\phi$ constructed above is satisfiable by an assignment of Hamming weight $k'$ if and only if $M$ accepts the empty string in $k$ steps. As BINARY NDTM HALTING$(k)$ can be solved in $2^k \cdot n^{O(1)}$ time, we may assume that $\log n \leq k$, since otherwise the reduction is trivial. Thus, letting $n'$ denote the number of variables of $\phi$, we have $k' \log n' = k^{O(1)}$ as needed, and the above construction is indeed a PPT. □

The remaining problems in Theorem 3 for which we need to show completeness are HITTING SET$(m)$, SET COVER$(n)$, and EXACT SET COVER$(n)$. Since it is well known that HITTING SET$(m) \equiv_{ppt}$ SET COVER$(n)$ and EXACT HITTING SET$(m) \equiv_{ppt}$ EXACT SET COVER$(n)$, we finish the proof of Theorem 3 by showing WK[1]-completeness for HITTING SET$(m)$.

**Lemma 14.** HITTING SET$(m)$ *is* WK[1]-*complete.*

*Proof.* HITTING SET$(m)$ can be shown to be in WK[1] by a similar argument used in Lemma 12. To show WK[1]-hardness, we reduce from EXACT HITTING SET$(m)$. Let $(U, \mathcal{F})$ be an input instance to EXACT HITTING SET$(m)$, with $\mathcal{F} = \{F_1, \ldots, F_m\}$ and $U := [n]$. We can assume $\log n \leq k$ by identifying identical vertices. Let $V = \bigcup_{i \in [m]} \{v_{i,j} : j \in F_i\}$; this will be the vertex set of our HITTING SET$(m)$ instance. To ensure consistency between the selections in different sets, we consider the binary expansions of the vertices in $U$. Let $i, i' \in [m]$, $i \neq i'$, and $\ell \in [\log n]$. Let $A_{(i,i',\ell)}$ consist of vertices $v_{i,j}$ for every $j \in F_i \cap F_{i'}$ such that the $\ell$'th bit of the binary expansion of $j$ is 0, and $B_{(i,i',\ell)}$ consist of vertices $v_{i',j}$ for every $j \in F_i \cap F_{i'}$ such that the $\ell$'th bit of the binary expansion of $j$ is 1. Let $C_{(i,i')} = \{v_{i,j} : j \in F_i \setminus F_{i'}\}$. We add the edge $E_{(i,i',\ell)} = A_{(i,i',\ell)} \cup B_{(i,i',\ell)} \cup C_{(i,'i)}$ to the HITTING SET$(m)$ instance for all $i, i' \in [m]$, $i \neq i'$, and $\ell \in [\log n]$. Finally, for every $i \in [m]$, we add the edge $E_i = \{v_{i,j} : j \in F_i\}$ to our output instance.

Let $\mathcal{E}$ denote the resulting edge set, and set $k = m$. If $(U, \mathcal{F})$ has an exact hitting set, then this immediately gives a hitting set for $(V, \mathcal{E})$ of size $m$: For every pair of edges $F_i$ and $F_{i'}$, either their intersection is hit, in which case exactly one of $A_{(i,i',\ell)}$ and $B_{(i,i',\ell)}$ is hit for every $\ell$, or the symmetric difference is hit in both sets, in which case $C_{(i,'i)}$ is hit. On the other hand, assume that $(V, \mathcal{E})$ has a hitting set of size $m$; then the edges $E_i$ ensure that this hitting set corresponds to selecting one vertex from each edge $F_i \in \mathcal{F}$. We argue that these vertices must form an exact hitting set for $(U, \mathcal{F})$. Consider thus a pair of sets $F_i$ and $F_{i'}$. Clearly, one vertex is selected in each edge. If both selected vertices are in $F_i \cap F_{i'}$, but the vertices differ, then one of the edges $E_{(i,i',\ell)}$ or $E_{(i',i,\ell)}$ is not hit, where $\ell$ is a bit distinguishing the binary expansions of these two vertices. If the vertex selected from $F_i$ lies in $F_i \cap F_{i'}$, but the vertex selected from $F_{i'}$ lies in the symmetric difference of $F_i$ and $F_{i'}$, let $\ell$ be a bit equalling 1 in the binary expansion of the vertex selected from $F_i$ (this exists by construction). Then the edge $E_{(i,i',\ell)}$ is not hit: Clearly, $B_{(i,i',\ell)}$ and $C_{(i,i')}$ are not hit, and by choice neither is $A_{(i,i',\ell)}$. Thus, if the two selected vertices are two distinct vertices, both these vertices must lie in the symmetric difference of $F_i$ and $F_{i'}$. This implies that for every pair of sets, the vertices selected from these sets are either identical vertices in the intersection, or two distinct vertices in the symmetric difference. It follows that the set of all selected vertices forms a valid exact hitting set for $(U, \mathcal{F})$. □

### 4.2 Satisfiability Problems

We now turn our attention to two important satisfiability problems. The first is MIN ONES $d$-SAT$(k)$, which is the problem of determining whether a $\Gamma_{1,d}$-formula ($d$-CNF) has a satisfying assignment of weight at most $k$. This problem is similar to the class-defining $\Gamma_{1,d}$-WSAT$(k \log n)$, except that as the problem is FPT (see [32]), we can simply parameterize by $k$. As $d$ is a constant, this can express in polynomial space any type of constraint involving at most $d$ variables. For example, this implies that the $\mathcal{H}$-FREE GRAPH EDITING$(k)$ problem is in WK[1] for every fixed set of forbidden induced subgraphs $\mathcal{H}$ (see [12]). The second satisfiability problem we deal with is LOCAL CIRCUIT SAT($k \log n$) which defines the class VC$_1$ of the VC-hierarchy defined by Harnik and Naor [24].

**Theorem 4.** MIN ONES $d$-SAT($k$) *is* WK[1]-*complete for every* $d \geq 3$.

*Proof.* MIN ONES $d$-SAT($k$) is easily seen to be in WK[1] by a reduction to $\Gamma_{1,d}$-WSAT($k \log n$). Let $\phi$ be the input formula with $n$ variables. We note that we may assume $\log n \leq k$, or else solve the problem in time polynomial in $n$ using the $2^{O(k)}$-time FPT algorithm [32]. Thus MIN ONES $d$-SAT($k$) reduces to MIN ONES $d$-SAT($k \log n$) by PPT. Further, we can create a trivial formula which is satisfiable at any weight from 0 to $k$, *e.g.* by adding $k$ clauses, each containing two negative literals of variables not occurring in the rest of the formula. This gives the PPT to $\Gamma_{1,d}$-WSAT($k \log n$).

To prove WK[1]-hardness, we show a PPT from EXACT HITTING SET($m$) which is WK[1]-hard by Theorem 3. Let $(V, \mathcal{E})$ be an instance of EXACT HITTING SET with $|V| = n$ and $|\mathcal{E}| = m$. First, we show that using binary trees, we can build *selection formulas* of polynomial size that force at least one member of each edge $E \in \mathcal{E}$ to be selected to a solution, while requiring only $\log |E|$ true variables. Recall that we can assume $|E| \leq n \leq 2^m$ for each edge $E \in \mathcal{E}$. We construct a $\Gamma_{1,3}$ formula $\phi$ with $n' = \sum_{E \in \mathcal{E}} |E| + n$ variables as follows. For each edge $E \in \mathcal{E}$, our formula $\phi$ will contain $|E|$ variables, in additional to global variables associated with the vertex set $V$.

Let $E \in \mathcal{E}$ and $\ell = \lceil \log |E| \rceil$. Build first a binary tree with root $x_0$ and where node $x_i$ has children $x_{2i+1}$ and $x_{2i+2}$, up to depth $\log |E|$. Create a clause $(x_0)$, and for every node $x_i$ a clause $(x_i \to (x_{2i+1} \vee x_{2i+2}))$. It is not hard to see that this construction requires at least one node on every level of the binary tree to be true, and that it can be satisfied by setting the nodes along a path from the root to a leaf to be true. Next, for every leaf $x'$ of the tree, add a clause $(x' \to v)$ for some $v \in E$, forming a surjective mapping from the leafs to vertices in $E$. This finishes the selection gadget for $E$, adding a cost of $\ell + 1$ true variables among the $x_i$. All clauses used in the gadget contain at most 3 literals. Finally, we add a conjunction of clauses of size 2 with negative literals, to constrain the formula to contain exactly one variable corresponding to the elements of $E$, as required by the EXACT HITTING SET problem. Create selection and constraint gadgets for every edge in $\mathcal{E}$, and let $k' = \sum_{E \in \mathcal{E}} (\lceil \log |E| \rceil + 1) + m$. The resulting formula is satisfiable if and only if $(V, \mathcal{E})$ has an exact hitting set, and $k'$ is clearly an upper bound on the possible number of true variables in a solution. □

The input to the LOCAL CIRCUIT SAT($k \log n$) problem defined by Harnik and Naor [24] is a string of length $n$ and a circuit $C$ with $k + k \log n$ inputs, of size $k + k \log n$. An instance is positive if there are $k$ positions $i_1, \ldots, i_k$ in the string such that feeding the contents of the positions to the first $k$ inputs, and the binary expansions of $i_1, \ldots, i_k$ to the remaining inputs, causes the circuit to accept. As Harnik and Naor note, we may equivalently assume the circuit to have size polynomially bounded in $k \log n$, rather than exactly $k + k \log n$.

**Theorem 5.** LOCAL CIRCUIT SAT($k \log n$) *is complete for* WK[1].

*Proof.* To show WK[1]-hardness, we reduce from CLIQUE($k \log n$) which is WK[1]-hard by Theorem 3. Let an instance $(G, k)$ be given. Assume that $n = 2^\ell$, or else pad the instance with isolated vertices (at most doubling the size). The input to the circuit is the adjacency matrix written row by row as a string of length $n' = n^2$, modified to have a 1 in each diagonal entry (for ease of presentation). The circuit gets input from $k^2$ positions, with each position coded in $\log n' = 2 \log n$ bits; here $n = 2^\ell$ ensures that there is a trivial way of converting between matrix positions and places in the string. The first part of the string thus provides the circuit with $k^2$ entries of the adjacency matrix, and it simply checks that these are all ones. The second part contains the position and the circuit checks that all the positions can be obtained by taking all $k^2$ combinations of concatenating the numbers of two vertices (size $\log n$). For this we may fix any desired ordering of the positions, hence the checking comes down to hardwired equalities in an appropriate way (e.g. we might decide that the first $k$ positions correspond to the first vertex hence the first $\log n$ bit-positions of the first $k$ positions should be bitwise the same). It is easy to check the correspondence between a $k$-clique in $G$ and a choice of variable assignments such that the circuit accepts.

To show membership in WK[1], we reduce to BINARY NDTM HALTING($k$) which is WK[1]-complete according to Theorem 3. The number of steps will be polynomial in $k \log n$. It is well known that a Turing machine can be constructed to simulate a fixed circuit in a number of steps that is polynomial in the circuit size. To simulate an instance of LOCAL CIRCUIT SAT we additionally encode the input string of length $n$ into the Turing machine. The machine guesses the $k$ positions in $k \log n$ steps, writes the according contents of the string onto the tape, followed by the positions, and then simulates the circuit. □

### 4.3 Path and Cycle Problems

We next prove hardness and completeness results with respect to WK[1] for well studied path and cycle problems. In particular, we prove the following two theorems:

**Theorem 6.** *The* DISJOINT PATHS($k$) *and* DISJOINT CYCLES($k$) *problems are* WK[1]-*hard*.

**Theorem 7.** *The following problems are complete for* WK[1]:

- MULTICOLORED PATH($k$) *and* DIRECTED MULTICOLORED PATH($k$).
- MULTICOLORED CYCLE($k$) *and* DIRECTED MULTICOLORED CYCLE($k$).

We note that the "uncolored" versions of both problems in Theorem 7 are in WK[1], but we do not know whether they are complete or not. Nevertheless, the above four problems are interesting in their own right, and algorithms for them are used as subroutines in the classical color-coding algorithms for their uncolored counterparts [4].

To prove both theorems above we go through an intermediate problem known as the DISJOINT FACTORS($k$) problem, and mimic the construction used in [10] to show that this problem is WK[1]-complete. The input to DISJOINT FACTORS($k$) is a string $S$ of length $n$ over the alphabet $[k]$. The goal is to determine whether there exists a set of $k$ disjoint substrings $S_1, \ldots, S_k$ of $S$, where $S_i$ of the form $i \cdots i$ (i.e. a factor) for each $i \in [k]$. Bodlaender *et al.* [10] show that this problem is solvable in $2^{O(k)} \cdot n$ time, and thus is in EXPT.

**Lemma 15.** *The* DISJOINT FACTORS($k$) *problem is* WK[1]-*hard*.

*Proof.* We construct a PPT from MULTICOLORED HITTING SET($m$) which is WK[1]-hard according to Theorem 3. Given an instance $(V, \mathcal{E}, c, k)$ of MULTICOLORED HITTING SET($m$), with $|V| = n$, $|\mathcal{E}| = m$, and $c : V \to [k]$, we create a string $S$ of size polynomial in $n + \bigcup_{E \in \mathcal{E}} |E|$ as an instance of DISJOINTFACTORS as follows: Our alphabet will consist of one symbol $\mathbf{e}$ for every edge $e \in \mathcal{E}$, and of at most $k \cdot \lceil \log n \rceil$ auxiliary symbols: For every color $i \in [k]$, we have $\lceil \log |V_i| \rceil$ symbols $\mathbf{a}_1^i, \ldots \mathbf{a}_{\lceil \log |V_i| \rceil}^i$, where $V_i \subseteq V$ is the subset of vertices with $c(v) = i$. Clearly we can identify vertices which are included in the same set of edges, and therefore we can assume that $\log n = O(m)$, which makes our reduction a PPT. Our string $S$ will contain substrings corresponding to vertices; to a vertex $v \in V$ contained in the edges $E_1, \ldots E_\ell \in \mathcal{E}$, we assign the substring $S_v = \mathbf{e}_1 \mathbf{e}_1 \ldots \mathbf{e}_\ell \mathbf{e}_\ell$. The symbols corresponding to edges will appear only inside such substrings $S(v)$. For every color $i \in [k]$, we will build a "selection gadget" for choosing a vertex of the given color analogous to the one described in [10, Lemma 2]. The gadget will ensure that we are able to pick factors contained in $S(v)$ if and only if $v$ is the chosen vertex from its color class.

Let us describe the selection gadget for color $i \in [k]$ more precisely. We assume that the number of vertices $n_i$ of this color is a power of 2 (otherwise we can just add isolated vertices), and we let $V_i = \{v_1, \ldots, v_{n_i}\}$. If we had only two vertices $v_1$ and $v_2$ to pick from, we could implement the gadget in the following manner: $S[1,2] = \mathbf{a}_1^i S(v_1) \mathbf{a}_1^i S(v_2) \mathbf{a}_1^i$, where $\mathbf{a}_1^i$ is some symbol which does not appear inside $S(v_1)$ nor $S(v_2)$. By choosing the factor for $\mathbf{a}_1^i$, we prevent selecting factors from either $S(v_1)$ or $S(v_2)$, which corresponds to selecting whether to hit all sets which include $v_1$, or all sets which include $v_2$. We can apply this construction in a recursive manner; if the substring $S[1, 2^j]$ implements the selection of a vertex from $\{v_1, \ldots, v_{2^j}\}$, and if $S[2^j+1, 2^{j+1}]$ implements the selection of a vertex from $\{v_{2^j+1}, \ldots, v_{2^{j+1}}\}$, then the substring $\mathbf{a}_{j+1}^i S[1, 2^j] \mathbf{a}_{j+1}^i S[2^j+1, 2^{j+1}] \mathbf{a}_{j+1}^i$ selects a single vertex from $\{v_1, \ldots, v_{2^{i+1}}\}$ (note that we need only one auxiliary symbol per level of the recursion; symbols $\mathbf{a}_1^i, \ldots, \mathbf{a}_j^i$ appear inside both $S[1, 2^j]$ and $S[2^j+1, 2^{j+1}]$). It is easy to check that the only way of selecting factors for every edge symbol $\mathbf{e}$ is selecting in the described gadgets substrings $S(v)$ corresponding to vertices in a multicolored hitting set of size $k$ for $(V, \mathcal{E}, c)$. The lemma thus follows. □

Bodlaender *et al.* [10] provide polynomial parametric transformations from DISJOINT FACTORS($k$) to DISJOINT CYCLES($k$) and DISJOINT PATHS($k$). This, in combination with the lemma above provides the proof for Theorem 6. To prove Theorem 7, we show that DIRECTED MULTICOLORED PATH($k$) and MULTICOLORED PATH($k$) are WK[1]-complete. The corresponding cycle problems in Theorem 7 follow immediately from this, and are thus omitted.

**Lemma 16.** *The* DIRECTED MULTICOLORED PATH($k$) *is* WK[1]-*complete*.

*Proof.* It is easy to see that DIRECTED MULTICOLORED PATH($k$) $\in$ WK[1], by reducing this problem to NDTM HALTING($k \log n$). Let $G$ be a directed $k$-colored graph on $n$ vertices given as input to DIRECTED MULTICOLORED PATH($k$). First note that as DIRECTED MULTICOLORED PATH($k$) can be solved in $2^{O(k)} \cdot n^{O(1)}$ time [4], we can assume $\log n = O(k)$. We construct a Turing machine $M$ that encodes within its state-space the adjacency matrix of $G$. It is easy to see that such a Turing machine can be programmed to determine non-deterministically in $O(k)$ steps whether $G$ has a multicolored path on $k$ vertices, by guessing $k$ vertices of different colors in $G$, and then checking whether these vertices form a path. Since $|M| = O(n^2)$ and $\log n = O(k)$, this gives the desired PPT.

To show hardness, we reduce from DISJOINT FACTORS($k$). Given an input string $S$ to DISJOINT FACTORS($k$) over the alphabet $[k]$, we construct a directed $k$-colored graph $G$ which has a vertex corresponding to each factor of $S$. Each vertex is colored according to the starting (or ending) letter of its corresponding factor, and there is an edge $(u, v)$ in $G$ if the factor corresponding to $u$ is strictly to the left of the factor corresponding to $v$ in $S$. It is easy to verify that $G$ has a multicolored path of length $k$ iff $S$ has $k$ disjoint factors. Thus, DISJOINT FACTORS($k$) $\leq_{ppt}$ DIRECTED MULTICOLORED PATH($k$), and the lemma is proven. □

**Lemma 17.** MULTICOLORED PATH($k$) *is* WK[1]-*complete*.

*Proof.* The argument showing that MULTICOLORED PATH($k$) $\in$ WK[1] is very similar to the one used for DIRECTED MULTICOLORED PATH($k$). To show hardness, we provide a PPT from DIRECTED MULTICOLORED PATH($k$). Let $G$ be a directed $k$-colored graph on $n$ vertices given as input to DIRECTED MULTICOLORED PATH($k$). We construct a $k'$-colored graph $G'$ on $O(n)$ vertices with $k' = 3k+4$. First, we split each vertex $v$ of color $c$ in $G$ into three vertices: $v_{in}, v_{mid}$, and $v_{out}$, of colors $c_{in}, c_{mid}$ and $c_{out}$ respectively, forming a path $v_{in}, v_{mid}, v_{out}$ in $G'$. A directed edge $(u,v)$ in $G$ will be transformed to an edge $\{u_{out}, v_{in}\}$ in $G'$. We add to $G'$ four additional vertices $s_{in}, s_{out}, t_{in}$ and $t_{out}$, and assign to them four unique colors. The source vertex $s_{in}$ will be connected only to $s_{out}$, and analogously the sink vertex $t_{out}$ will be connected only to $t_{in}$. We connect $s_{out}$ to every vertex $v_{in}$, $v \in V(G)$, and we connect $t_{in}$ to every $v_{out}$, $v \in V(G)$.

Note that a multicolored path of length $k'$ in $G'$ must have $s_{in}$ and $t_{out}$ as its endpoints: These vertices are the only representatives of their color classes, and therefore have to appear on the path, and both have degree exactly one, which means that they can only be endpoints of the path. Another property of $G'$ is the fact that any multicolored path which visits one of the vertices $v_{in}, v_{mid}$ or $v_{out}$ has to visit the other two vertices as well in a consecutive manner, which can be proven by a straightforward case analysis. A path starting at $s_{in}$ will therefore always follow the directed edges of $G$ in the "right direction": When arriving at $v_in$ via some incoming edge it will proceed to $v_{mid}$ and $v_{out}$ and then take some outgoing edge. This observation shows that $G$ has a path of length $k$ iff $G'$ has a path of length $k'$. Since our construction is clearly a PPT, the lemma is proven. □

### 4.4 Further Problems

In this section we investigate WK[1]-completeness and hardness of various problems for which kernelization lower bounds were obtained by Dom, Lokshtanov, and Saurabh [17].

**Theorem 8.** *The following problems are complete for* WK[1]*:*

- CONNECTED VERTEX COVER$(k)$.
- CAPACITATED VERTEX COVER$(k)$.
- STEINER TREE$(k+t)$.
- SMALL SUBSET SUM$(k)$.
- UNIQUE COVERAGE$(k)$.

The first four problems in the theorem above all have polynomial parameteric transformations from HITTING SET$(m)$ [17, Theorems 2 and 7], and are thus WK[1]-hard by Lemma 14. To see that UNIQUE COVERAGE$(k)$ is also WK[1]-hard, consider the following easy PPT from EXACT HITTING SET$(m)$. Let $(V, \mathcal{E})$ be the input instance with $|\mathcal{E}| = m$, and recall that we can assume as usual that $\log n \leq m$. For each $v \in V$ make a set $F_v$ containing all sets $E \in \mathcal{E}$ with $v \in E$. It is easy to see that any set $\mathcal{F}' \subseteq \{F_v : v \in V\}$ which uniquely covers $k = m$ elements of $\mathcal{F}$ directly corresponds to an exact hitting set for $(V, \mathcal{E})$. Thus, all problems in Theorem 8 are WK[1]-hard, and so to complete the proof of theorem we show that all these problems are members of WK[1].

For all five problems, membership in WK[1] is shown by PPTs to NDTM HALTING$(k \log n)$. In all cases the input will be included in the description of the Turing machine to allow for a fast verification of guessed solutions. For STEINER TREE$(k+t)$ and SMALL SUBSET SUM$(k)$ the PPTs are straightforward. The PPT for UNIQUE COVERAGE$(k)$ is also straightforward once we realize that it sufficient to consider a solution of size at most $k$, and that the $O(4^k)$-size kernel of Moser et al. [33] lets us assume that the logarithm of the input size is polynomially bounded in $k$. We thus state the following lemma without proof.

**Lemma 18.** STEINER TREE$(k+t)$, SMALL SUBSET SUM$(k)$, *and* UNIQUE COVERAGE$(k)$ *all have* PPT*s to the* NDTM HALTING$(k \log n)$ *problem.*

For the two remaining vertex cover variants in Theorem 8 the reductions are a bit more subtle. Both of them utilize the so-called Buss rule used in the classical $O(k^2)$ kernel for VERTEX COVER$(k)$ [11]. This allows the output Turing machine in the reduction to quickly verify solutions that it guesses. More details are given in the proofs of the two lemmas below.

**Lemma 19.** CONNECTED VERTEX COVER($k$) $\leq_{ppt}$ NDTM HALTING($k \log n$).

*Proof.* Given an instance $(G, k)$ of CONNECTED VERTEX COVER($k$), we first identify the set $T$ of all vertices of degree greater than $k$, and output a Turing machine that never halts if $|T| > k$ since the vertices of $T$ must be in every vertex cover of size $k$ for $G$. The remaining budget $k' = k - |T|$ must be spend on a set $N$ of vertices which covers all edges not incident on $T$ and such that $G[T \cup N]$ is connected. Since all vertices outside $T$ have degree at most $k$ there can be at most $\binom{k}{2}$ uncovered edges, or we may output a machine that never halts (as vertices in $N$ cover at most $k|N|$ edges altogether). We construct a Turing machine $M$ by encoding in its state-space the set of at most $\binom{k}{2}$ uncovered edges, the set $T$, the graph $G$, and the budget $k'$. It is now not difficult to see that by properly programming $M$, the machine will determine in $k^{O(1)}$ non-deterministic steps whether $G$ has a connected vertex cover of size $k$. □

**Lemma 20.** CAPACITATED VERTEX COVER($k$) $\leq_{ppt}$ NDTM HALTING($k \log n$).

*Proof.* Let $(G, \alpha, k)$ be an instance of CAPACITATED VERTEX COVER($k$), where $G = (V, E)$ is a graph on $n$ vertices. We again identify the set $T$ of vertices with degree exceeding $k$ in $G$, and output a non-halting machine in case $|T| > k$. We also output a non-halting machine if the capacity of some vertex in $T$ is lower than the number of its incident edges by more than $k$; conversely we may delete a vertex of $T$ if its capacity suffices for all incident edges (decreasing $k$ by one). By selecting a set $N$ of $k - |T|$ further vertices at most $k^2$ edges can be covered; this includes edges not incident with $T$ but also edges incident with some vertex of $T$ for which the budget does not suffice. The output Turing machine $M$ is hardwired with an encoding of the set $T$ along with the input graph, including the degrees and capacity of all vertices, to essentially allow random access to all these values. The machine $M$ proceeds as follows:

1. Guess a set $N$ of at most $k' = k - |T|$ vertices. Verify that all edges have at least one endpoint in $T \cup N$ (this only has to be done for the at most $k^2$ edges not incident to $T$).
2. For every edge with both endpoints in $T \cup N$, guesses the vertex which covers it. Register on the tape for every vertex in $T \cup N$ the number of incident edges that the vertex does not have to cover (due to it being covered at its other endpoint).
3. Verify for every vertex in $T \cup N$ that its capacity suffices to cover all remaining edges, that is, that its capacity is at least its degree minus the number of edges covered by another vertex.

It can be verified that the required number of steps can be bounded by a polynomial in $k$ and that the total machine size and construction time are polynomial in $n$. □

## 5 Problems in Higher Levels

In this section we investigate the second level of the MK- and WK-hierarchies, and present some complete and hard problems for these classes.

### 5.1 MK[2]

According to Theorem 1, MK[2] is the PPT-closure of the classical CNF satisfiability problem where the parameter is taken to be the number of variables in the input formula. The PPT-equivalence of this problem to HITTING SET($n$) and SET COVER($m$) is well known.

**Theorem 9.** HITTING SET($n$) *and* SET COVER($m$) *are complete for* MK[2].

Heggernes et al. [25] consider the problems RESTRICTED PERFECT DELETION($|X|$) and RESTRICTED WEAKLY CHORDAL DELETION($|X|$), where the input is a graph $G$, a set of $\ell$ vertices $X$ of $G$ such that $G - X$ is perfect (respectively weakly chordal), and an integer $k$, and the task is to select at most $k$ vertices $S \subseteq X$ such that $G - S$ is perfect (respectively weakly chordal). The following corollary is immediate from Theorem 9 and PPTs given in [25].

**Corollary 1.** RESTRICTED PERFECT DELETION($\ell$) *and* RESTRICTED WEAKLY CHORDAL DELETION($\ell$) *are hard for* MK[2].

## 5.2 WK[2]

The following theorem establishes WK[2]-completeness for the following reparameterizations of well-known W[2]-complete problems.

**Theorem 10.** *The following problems are complete for* WK[2]:

- HITTING SET($k \log n$) *and* SET COVER($k \log m$).
- DOMINATING SET($k \log n$) *and* INDEPENDENT DOMINATING SET($k \log n$).
- STEINER TREE($k \log n$)

From Theorem 10, we immediately get the following corollary via PPTs by Lokshtanov [31] and Heggernes *et al.* [25].

**Corollary 2.** *The following problems are all hard for* WK[2]:

- WHEEL-FREE DELETION($k \log n$).
- PERFECT DELETION($k \log n$).
- WEAKLY CHORDAL DELETION($k \log n$).

For the first four problems in Theorem 10, the results follow easily. The PPT-equivalence between $\Gamma_{2,1}$-WSAT($k \log n$), HITTING SET($k \log n$), SET COVER($k \log m$), and DOMINATING SET($k \log n$) are well known, and for INDEPENDENT DOMINATING SET($k \log n$), a PPT to $\Gamma_{2,1}$-WSAT($k \log n$) is trivial and a PPT from DOMINATING SET($k \log n$) can be produced by standard methods.

The story is different with STEINER TREE($k \log n$). While WK[2]-hardness for this problem follows immediately from *e.g.* the PPT from HITTING SET($k \log n$) given in [17], showing membership in WK[2] is more challenging. To facilitate this and other non-trivial membership proofs, we consider the issue of a machine characterization of WK[2], similarly to the WK[1]-complete BINARY NDTM HALTING($k$) problem. The natural candidate would be MULTI-TAPE NDTM HALTING($k \log n$), as this same problem with parameter $k$ is W[2]-complete [14]. However, while the problem with parameter $k \log n$ is easily shown to be WK[2]-hard, we were so far unable to show WK[2]-membership. On the other hand, we define the following extension of a single-tape non-deterministic Turing machine problem which is WK[2]-complete. We name the corresponding halting problem NDTM HALTING WITH FLAGS.

**Definition 5.** *A* (single-tape, non-deterministic) *Turing machine with flags is a standard* (single-tape, non-deterministic) *Turing machine which in addition to its working tape has access to a set $F$ of* flags. *Each state transition of the Turing machine has the ability to read and/or write a subset of the flags. A transition that* reads *a set $S \subseteq F$ of flags is only applicable if all flags in $S$ are set. A transition that* writes *a set $S \subseteq F$ of flags causes every flag in $S$ to be set. In the initial state, all flags are unset. Note that there is no operation to reset a flag.*

**Theorem 11.** NDTM HALTING WITH FLAGS($k \log n$) *is* WK[2]-*complete.*

*Proof.* Showing WK[2]-hardness is easy by reduction from HITTING SET($k \log n$). In fact, the hitting set instance can be coded directly into the flags, without any motion of the tape head – simply construct a machine that non-deterministically makes $k$ non-writing transitions, each corresponding to including a vertex in the hitting set, followed by one verification step. The machine has $m$ flags, one for every set in the instance, and a step corresponding to selecting a vertex $v$ activates all flags corresponding to sets containing $v$. Finally, the step to the accepting state may only be taken if all flags are set. By assuming $\log m \leq k \log n$ (or else solving the instance exactly) we get a PPT.

Showing membership in WK[2] can be done by translation to $\Gamma_{2,1}$-WSAT($k \log n$). The transition is similar to that in Lemma 13. The only complication is to enforce consistency of transitions which read and write sets of flags, but this is easily handled. Let $M_{e,t}$ signify that step number $t$ of the machine follows edge $e$ of the state diagram (as in Lemma 13). If transition $e$ has a flag $f$ as a precondition, then we simply add a clause

$$(\neg M_{e,t} \vee M_{e_{i_1},1} \vee \ldots \vee M_{e_{i_m},1} \vee \ldots \vee M_{e_{i_1},t-1} \vee \ldots \vee M_{e_{i_m},t-1}),$$

where $e_{i_1}, \ldots, e_{i_m}$ is an enumeration of all transitions in the state diagram which set flag $f$. The rest of the reduction proceeds without difficulty. □

**Lemma 21.** STEINER TREE($k \log n$) *is* WK[2]-*complete.*

*Proof.* As mentioned above, WK[2]-hardness for STEINER TREE($k \log n$) follows from the PPT from HITTING SET($k \log n$) given in [17]. We show membership in WK[2] by a reduction to the Turing machine problem with flags. Let $(G, T, k)$ be an instance of STEINER TREE. We make the following observations.

1. If two terminals $t, t' \in T$ are neighbors in $G$, they may be identified.
2. Assume that no two terminals are neighbors in $G$. Let $G'$ be $G$ with $N(t)$ replaced by a clique for every $t \in T$. Then, for any solution $S$, the graph $G'[S]$ must be connected.

In fact, a set $S$ is a Steiner tree if and only if $G'[S]$ is connected and every $t \in T$ is neighbor to a vertex in $S$. Thus, perform the reduction to $G'$ as described. The Turing machine then goes in two phases. First, it guesses the solution $S$ consisting of $k$ vertices, and checks (in poly($k$) time) the connectivity of $G'[S]$. Second, using the flags as in Theorem 11, it goes through the vertices of $S$ and verifies that every terminal is neighbor to at least one vertex. It accepts if both tests pass. □

## 6 Discussion

We have defined a hierarchy of classes of inefficient kernelization, akin to the M- and W-hierarchy of parameterized intractability. The fundamental distinction in the new hierarchy is between problems admitting polynomial Turing kernels (Turing-PK) and WK[1]-hard problems. This distinction does not seem to be addressable by previous lower-bound frameworks, as evidenced by problems such as LEAF OUT BRANCHING($k$) and CLIQUE($\Delta$), which admit polynomial Turing kernels but no standard polynomial kernels unless the polynomial hierarchy collapses. Thus, WK[1] $\nsubseteq$ Turing-PK is a new conjecture in kernelization theory. We showed that many natural parameterized problems to which the kernelization lower bounds apply are WK[1]-complete, indicating that the class is natural. Of course, our examples provide only a partial image of the WK[1] landscape. For example, the various kernelizability dichotomies that have been shown for CSP problems [29, 30] can be shown to imply dichotomies between problems with polynomial kernels and WK[1]-complete problems (and in some cases the third class of W[1]-hard problems).

Still, several questions remain. One is the WK[1]-hardness of PATH($k$) and CYCLE($k$); for these problems, we have only lower bound proofs in the framework of Bodlaender *et al.* [6], leaving the question of Turing kernels open. There are also several problems, including the work on structural graph parameters by Bodlaender, Jansen, and Kratsch [9, 26, 27], which we have not investigated. It is also unknown whether MULTI-TAPE NDTM HALTING($k \log n$) is in WK[2]. Furthermore, it would be interesting to know some natural parameterized problems which are WK[2]-complete under a standard parameter (e.g., $k$ rather than $k \log n$). On the more structural side, we have noted several connections between our hierarchy and previous hierarchies. It would be interesting if these could be made stronger. In particular, we would like to know if there are (classical or parameterized) complexity theoretical implications of polynomial Turing kernelizations for WK[1].

# A   Problem Zoo

Below we provide problem statements to all problems discussed in the paper. We adopt the notation that appends brackets at the end of problem names to specify the parameterization used for the specific problem. For instance, CONNECTED VERTEX COVER($k$) denotes the CONNECTED VERTEX COVER problem parameterized by the number $k$ of vertices in the solution.

$\Phi$-SAT:
**Input:** A formula $\phi \in \Phi$ with $n$ variables, and an integer $k$.
**Task:** Decide whether $\phi$ is satisfiable.

$\Phi$-WSAT:
**Input:** A formula $\phi \in \Phi$ with $n$ variables, and an integer $k$.
**Task:** Decide whether $\phi$ is satisfiable by an assignment of Hamming weight $k$ (an assignment that assigns exactly $k$ variables the boolean value 1).

BINARY NDTM HALTING:
**Input:** A Turing machine $M$ of size $n$ with a binary alphabet, and an integer $k$.
**Task:** Decide whether $M$ halts on the empty string in $k$ steps.

CLIQUE:
**Input:** A graph $G$ with $n$ vertices, and an integer $k$.
**Task:** Decide whether $G$ has a clique of size $k$ (a pairwise adjacent subset of $k$ vertices).

CAPACITATED VERTEX COVER:
**Input:** A graph $G$ with $n$ vertices, a capacity function $\alpha : V(G) \to \mathbb{N}$, and an integer $k$.
**Task:** Decide whether $G$ has a capacitated vertex cover of size $k$ (a subset of $k$ vertices $S$ that are incident with each edge $G$, and such that each vertex $v \in S$ is incident with at most $\alpha(v)$ edges).

CONNECTED VERTEX COVER:
**Input:** A graph $G$ with $n$ vertices, and an integer $k$.
**Task:** Decide whether $G$ has a connected vertex cover of size $k$ (a connected subset of $k$ vertices $S$ that are incident with each edge of $G$).

DIRECTED MULTICOLORED CYCLE:
**Input:** A directed graph $G$, a coloring function $c : V \to [k]$, and an integer $k$.
**Task:** Decide whether $G$ has a multicolored directed cycle of length $k$ (a directed cycle which includes exactly one vertex from each color).

DIRECTED MULTICOLORED PATH:
**Input:** A directed graph $G$, a coloring function $c : V \to [k]$, and an integer $k$.
**Task:** Decide whether $G$ has a multicolored directed path of length $k$ (a directed path which includes exactly one vertex from each color).

DISJOINT CYCLES:
**Input:** A graph $G$ with $n$ vertices, and an integer $k$.
**Task:** Decide whether $G$ contains $k$ pairwise disjoint cycles.

DISJOINT FACTORS:
**Input:** A $n$-character string $S$ over the alphabet $[k]$.
**Task:** Decide whether there exists a set of $k$ non-overlapping substrings $S_1, \ldots, S_k$ of $S$ such that $S_i$ is of the form $i \cdots i$ for every alphabet symbol $i \in [k]$.

DISJOINT PATHS:
**Input:** A graph $G$ with $n$ vertices, and $k$ pairs of vertices $(s_1, t_1), \ldots, (s_k, t_k)$.
**Task:** Decide whether $G$ contains $k$ pairwise disjoint paths connecting $s_i$ to $t_i$ for all $i \in [k]$.

DOMINATING SET:
**Input:** A graph $G$ with $n$ vertices, and an integer $k$.
**Task:** Decide whether $G$ has a dominating set of size $k$ (a set $D$ of $k$ vertices for which every vertex not in $D$ has a neighbor in $D$).

EXACT HITTING SET:
**Input:** A hypergraph $(V, \mathcal{E})$ with $|V| = n$ and $|\mathcal{E}| = m$.
**Task:** Decide whether $(V, \mathcal{E})$ has an exact hitting set (a subset $S \subseteq V$ such that $|S \cap E| = 1$ for all $E \in \mathcal{E}$).

EXACT SET COVER:
**Input:** A hypergraph $(V, \mathcal{E})$ with $|V| = n$ and $|\mathcal{E}| = m$.
**Task:** Decide whether $(V, \mathcal{E})$ has an exact set cover (a subset $\mathcal{S} \subseteq \mathcal{E}$ of pairwise disjoint edges with $\bigcup \mathcal{S} = V$).

INDEPENDENT SET:
**Input:** A graph $G$ with $n$ vertices, and an integer $k$.
**Task:** Decide whether $G$ has an independent set of size $k$ (a pairwise non-adjacent subset of $k$ vertices).

HITTING SET:
**Input:** A hypergraph $(V, \mathcal{E})$ with $|V| = n$ and $|\mathcal{E}| = m$, and an integer $k$.
**Task:** Decide whether $G$ has a hitting set of size $k$ (a subset $S \subseteq V$ of size $k$ with $S \cap E \neq \emptyset$ for all $E \in \mathcal{E}$).

LOCAL CIRCUIT SAT:
**Input:** A circuit $C$ over $k + k \log m$ variables and of size $k + k \log m$, and a string $S$.
**Task:** Decide whether there is a list of $k$ positions $i_1, \ldots, i_k$ in $S$ such that feeding the contents of the positions to the first $k$ inputs, and the binary expansions of $i_1, \ldots, i_k$ to the remaining inputs, causes $C$ to accept.

MIN ONES $d$-SAT:
**Input:** A formula $\phi \in \Gamma_{1,d}$ with $n$ variables, and an integer $k$.
**Task:** Decide whether $\phi$ is satisfiable by an assignment of Hamming weight at most $k$.

MULTICOLORED $\Phi$-WSAT:
**Input:** A formula $\phi \in \Phi$ over a variable set $X$ of size $n$, a coloring function $c : X \to [k]$, and an integer $k$.
**Task:** Decide whether $\phi$ is satisfiable by an multicolored assignment of Hamming weight $k$ (an assignment where no two variables of same color are assigned a 1).

MULTICOLORED CLIQUE:
**Input:** A graph $G = (V, E)$ with $|V| = n$, a coloring function $c : V \to [k]$, and an integer $k$.
**Task:** Decide whether $G$ has a multicolored clique of size $k$ (a clique containing exactly one vertex of each color).

MULTICOLORED CYCLE:
**Input:** A graph $G$, a coloring function $c : V \to [k]$, and an integer $k$.
**Task:** Decide whether $G$ has a multicolored cycle of length $k$ (a cycle which includes exactly one vertex from each color).

MULTICOLORED HITTING SET:
**Input:** A hypergraph $(V, \mathcal{E})$ with $|V| = n$ and $|\mathcal{E}| = m$, a coloring function $c : V \to [k]$, and an integer $k$.
**Task:** Decide whether $G$ has a multicolored hitting set of size $k$ (a hitting set which includes exactly one vertex from each color).

MULTICOLORED PATH:
**Input:** A graph $G$, a coloring function $c : V \to [k]$, and an integer $k$.
**Task:** Decide whether $G$ has a multicolored path of length $k$ (a path which includes exactly one vertex from each color).

NDTM HALTING:
**Input:** A Turing machine $M$ of size $n$, and an integer $k$.
**Task:** Decide whether $M$ halts on the empty string in $k$ steps.

PERFECT DELETION:
**Input:** A graph $G$ on $n$ vertices, and an integer $k$.
**Task:** Decide whether $G$ has at most $k$ vertices $S$ such that $G - S$ is perfect.

RESTRICTED PERFECT DELETION:
**Input:** A graph $G$ on $n$ vertices, a set of $\ell$ vertices $X$ of $G$ such that $G - X$ is perfect, and an integer $k$.
**Task:** Decide whether $G$ has at most $k$ vertices $S \subseteq X$ such that $G - S$ is perfect.

Restricted Weakly Chordal Deletion:
**Input:** A graph $G$ on $n$ vertices, a set of $\ell$ vertices $X$ of $G$ such that $G - X$ is weakly chordal, and an integer $k$.
**Task:** Decide whether $G$ has at most $k$ vertices $S \subseteq X$ such that $G - S$ is weakly chordal.

Set Cover:
**Input:** A hypergraph $(V, \mathcal{E})$ with $|V| = n$, $|\mathcal{E}| = m$, and $\max_{E \in \mathcal{E}} |E| = d$. Also, an integer $k$.
**Task:** Decide whether $(V, \mathcal{E})$ has a set cover of size $k$ (a subset $\mathcal{S} \subseteq \mathcal{E}$ of $k$ edges with $\bigcup \mathcal{S} = V$).

Small Subset Sum:
**Input:** An integer $k$, a set $S$ of integers of size at most $2^k$, and an integer $t$.
**Task:** Decide whether there are at most $k$ distinct integers in $S$ that sum up to $t$.

Steiner Tree:
**Input:** A graph $G = (V, E)$ with $|V| = n$, a set of $t$ terminals $T \subseteq V$, a set of $\ell$ non-terminals $N \subseteq V$, and an integer $k$.
**Task:** Decide whether there is a subset of at most $k$ non-terminals $N' \subseteq N$ such that $G[T \cup N']$ is connected.

Unique Coverage:
**Input:** A hypergraph $(V, \mathcal{E})$ with $|V| = n$ and $|\mathcal{E}| = m$, and an integer $k$.
**Task:** Decide whether there exists a subset $\mathcal{E}' \subseteq E$ such that at least $k$ vertices are contained in exactly one edge in $\mathcal{E}'$.

Weakly Chordal Deletion:
**Input:** A graph $G$ on $n$ vertices, and an integer $k$.
**Task:** Decide whether $G$ has at most $k$ vertices $S$ such that $G - S$ is weakly chordal.